\documentstyle{article}\begin{document}\centerline{\bf A non-linear equation from photopolymerization kinetics}\vskip .3in\centerline{M. L. Glasser   }

\centerline{Department of Physics and Center for Quantum Device Technology}

\centerline{Clarkson University}

\centerline{Potsdam, NY 13699-5820 (USA)}\vskip 1in
\centerline{ABSTRACT}

\begin{quote} In a medium where a photoreactive species produces N equally photoreactive radicals, by combining the Lambert-Beer law for the decay of light intensity, as a beam propagates into the medium, with the kinetic equation, one obtains a generalization of the Wegscheider equation from photobleaching theory. It is shown that this equation can be solved exactly, but implicitly, and can be reduced to a first order ordinary differential equation in a single reduced variable. \end{quote}

\newpage

\centerline{\bf Introduction}\vskip .1in

As a light beam propagates into a medium containing a various photoreactive species the beam intensity decreases with depth due to absorption proportional to the concentration of the active molecules as is described by the Lambert-Beer law [1].  By combining this law with the kinetic equation one obtains a non-linear first order integro-partial differential equation. Situations where the N radicals released lead to desirable effects, such as bleaching and polymerization,  have been studied for many years. An extensive list of references is presented in [2]. For the case of a single species the Lambert-Beer law for the light intensity
$$I(x,t)=I_0\exp[-\alpha\int_0^xC(u,t)du]\eqno(1)$$
combined with the kinetic equation
$$\frac{\partial C(x,t)}{\partial t}=-\phi\alpha I(x,t) C(x,t)\eqno(2)$$
gives Wegscheider's equation [3]
$$\frac{\partial C(x,t)}{\partial t}=-\phi\alpha I_0\exp[-\alpha\int_0^xC(u,t)du]C(x,t)$$
$$C(x,0)=C_0.\eqno(3)$$
whose solution has been known for many years [2]. Here $C$ is the concentration,  $\alpha$ is the absorption coefficient, and $\phi$ the quantum yield.  For the case of a slab $0<x<L$, in terms of reduced variables $T=\phi\alpha t$, $z=x/L$, $ S_0(z,T)=C(x,t)/C_0$, $\gamma=\alpha C_0L$,
the solution is 
$$S_0(z.T)=[1+e^{-\gamma z}(e^T-1)]^{-1}\eqno(4)$$
which is examined thoroughly in [2].

Systems are presently under investigation [4] where, not only the initial species, whose concentration is $C$, but its $N$ reaction products all have the same absorption coefficient. A similar analysis leads to the more general equation, expressed in terms of the  dimensionless variables given above,
$$\frac{\partial C(z,T)}{\partial T}=-e^{-N\gamma z}\exp[\gamma\int_0^zC(u,T)du]C(z,T)$$
$$C(z,0)=1.\eqno(5)$$
The aim of this note is to present the exact, albeit implicit, solution to (5).
\newpage

\vskip .2in\centerline{\bf Calculation}\vskip .1in

We first introduce the cumulative concentration $\sigma(z,T)=\int_0^zS(u,T)du$ for which $\sigma(z,0)=z$ and $\sigma(0,T)=0$. Then by integrating both sides of (5) over $z$ and then differentiating with respect to $z$, one finds
$$\frac{\partial^2}{\partial z\partial T}[\sigma(z,T)-Nz]=-e^{-\gamma[\sigma(z,T)-Nz]}\frac{\partial}{\partial z}[\sigma(z,T)-Nz].\eqno(6)$$
That is, for $f(z,T)=\sigma(z,T)-Nz$ one has the partial differential equation
$$\frac{\partial^2f(z,T)}{\partial z\partial T}+\left(\frac{\partial f}{\partial z}+N\right)e^{\gamma f(z,T)}=0\eqno(7)$$
with $f(z,0)=-(N-1)z$, $f(0,T)=0$.

Next, we introduce
$V(x,T)=-\log[f_z+N]$ to obtain
$$\frac{\partial^2 V(z,T)}{\partial z\partial T}=-\gamma\frac{\partial}{\partial t}(e^{-V(z,T)}+NV)\eqno(8)$$
which, after integration with respect to $T$ becomes
$$\frac{\partial V(z,T)}{\partial z}=\gamma(1-NV-e^{-V})$$
$$V(0,T)=T.\eqno(9)$$
From (9) we get the implicit relation
$$\int_T^{V(z,T)}\frac{ds}{1-Ns-e^{-s}}-\gamma z=0\eqno(10)$$
or, since $S=e^{-V}$, following a simple change of integration variable,
$$\int_{S(z,T)}^{e^{-T}}\frac{du}{u(1+N\log\; u-u)}-\gamma z=0.\eqno(11)$$

Finally, in terms of the new variables
$$\tau=\int_{\ln\; 2}^T\frac{du}{1-Nu-e^{-u}}$$
$$\xi=\gamma z-\tau\eqno(12)$$
$$S(z,T)=S(\xi),$$
we find that the solution to (5) has the implicit representation
$$\int_{1/2}^{S(\xi)}\frac{du}{u(1+N\ln\; u-u)}=\xi.\eqno(13)$$
\newpage

\vskip .2in\centerline{\bf Discussion}\vskip .1in

Let us first look at the case $N=0$, which renders (5) equivalent (under $S=-S_0 $) (mathematically, but not physically) to the Wegscheider equation [1].
By explicit integration
$\tau=\ln(e^T-1)$, $S(\xi)=e^{\xi}(e^{\xi}+1)^{-1}$ and we recover the solution (4).  Note that (13) is equivalent to the first order ordinary differential equation
$$S'(\xi)=\xi(1+N\ln\;\xi-\xi)\eqno(14)$$
subject to an appropriate initial condition. Eq. (14) should be useful for obtaining series approximations to
$S(z,T).$  It also indicates that for $N>0$ $S(z,T)$ is nonanalytic along the trajectory $\gamma z=\tau$.

\vskip .5in\noindent
{\bf Acknowledgements} The author is grateful to Mr.Venkata  Nekkanti for introducing him to this problem and thanks George Lamb and Chris Cosgrove for mathematical  suggestions.

\newpage \centerline{\bf References}\vskip .1in\noindent [1]  J.G. Calvert and J.N. Pitts, {\it Photochemistry}[ Wiley, NY (1966)]

\noindent [2] Guillermo Terrones and Arne J. Pearlstein, Macromolecules {\bf{34}}, 3195 (2001).

\noindent [3]  R. Wegscheider, Z. Phys. Chem.{\bf{103}}, 273 (1923).

\noindent [4]  V. Nekkanti (Private communication).

\end{document}